\documentstyle[osa,manuscript]{revtex}

\begin{document}
%\textwidth35pc
%\def\btt#1{{\tt$\backslash$#1}}
%\begin{document}
%\draft
%\preprint{~~}

%\font\one=cmssi12.tfm
%\font\Bxfnt=cmsy10 scaled\magstep1
%\def\BoxOp{\mathord{\hbox{\Bxfnt\char116 }\llap{\Bxfnt\char117 }}}

\def\s{\smallskip}
\def\n{\noindent}
\def\m{\medskip}
\def\b{\bigskip}
\def\btt#1{{\tt$\backslash$#1}}
\def\fr#1#2{\mbox{$\frac{#1}{#2}$}}
  \def\n{\noindent}
  \def\ve{{\varepsilon}}
  \def\th{\theta}
  \def\a{\alpha}
  \def\b{\beta}
\def\d{d}
  \def\be{\begin{equation}}
  \def\ee{\end{equation}}
  \def\bq{\begin{eqnarray}}
  \def\eq{\end{eqnarray}}
  \def\p{\partial}
  \def\({\left(}
  \def\){\right)}
  \def\lb{\left[}
  \def\rb{\right]}

\title{Oscillating non-singular relativistic spherical model}

\author{Naresh Dadhich\thanks{E-mail: nkd@iucaa.ernet.in}} 
\address{Inter--University Center for Astronomy and Astrophysics, 
Post Bag 4, Ganeshkind, Pune 411007, India}

\author{A.K. Raychaudhuri}
\address{Relativity and Cosmology Research Centre, Jadavpur 
University, Calcutta 700 032, India}

\date{\today}
\maketitle
\begin{abstract}
 A particular choice of the time function in the recently presented 
spherical solution by Dadhich [1] leads to a singularity free 
cosmological model which 
oscillates between two regular states. The energy-stress tensor involves 
anisotropic pressure and a heat flux term but is consistent with the 
usual energy conditions (strong, weak and dominant). By choosing the 
parameters suitably one can make the model consistent with observational 
data. An interesting feature of the model is that it 
involves blue shifts as in the quasi steady state model [2] but without 
violating general relativity.
\end{abstract}

%\end{verbatim}
\newpage
%\baselineskip = 2 \baselineskip  % double space the text

\n Following the discovery of non-singular cylindrically symmetric perfect 
fluid exact cosmological solution of the Einstein equation by Senovilla [3], 
some spherically  
symmetric non-singular models have been presented by Dadhich et.al [1,4]. 
These models have an energy-stress tensor with anisotropic pressure and 
heat flux but obeying the strong, weak and dominant energy conditions. 
The metric has a time function which can be arbitrarily chosen subject to 
the constraint of non-singularity and the energy conditions. It turns out 
that there exist different such choices which will be discussed in a 
detailed paper separately. In this letter we shall confine to the  
choice that gives an oscillatory behaviour of the universe without any 
singularity. \\

\n The authors are not aware of any oscillatory singularity free model in 
classical general relativity (GR) while there are some oscillatory 
behaviour models proposed in the recent formulation of of quasi steady 
state cosmology (QSSC) [2]. As the QSSC models predict the possibility of 
blue shift, our model would also admit that possibility. Thus should 
observations in future reveal blue shifts, it may simply indicate that 
the matter in the uinverse is not perfect fluid and one need not bring in 
the ideas of non-conservation as in QSSC contradicting GR. \\

\n Our oscillatory model is described by the metric [1],  
\be
\d s^2 = (r^2 + P) \d t^2 - \frac{2r^2 + P}{r^2 + P} \d r^2 - r^2 
(\d\th^2 + sin^{2}\th \d \varphi^2)  \\ 
\ee

\n where $P = P(t)$ which can be chosen arbitrarily. The choice $P(t) 
= a^2 + b^2 cos{\omega}t$ with $a^2 > b^2$ will render oscillatory 
behaviour to the model without encountering divergence of any 
kinematical and physical parameters. In Ref. [1]  the choice made was 
$P(t) = a^2 + b^2t^2$, which of course did not give oscillatory behaviour. \\ 

\n The energy-stress tensor for imperfect fluid is given by [5],

\be
T_{ik} = (\rho + p)u_iu_k - pg_{ik} + {\bigtriangleup}p[c_i c_k +
\frac{1}{3}(g_{ik} - u_iu_k)] + 2qc_{(i}c_{k)} \\
\ee

\n where $u_i$ and $c_i$ are respectively unit timelike and spacelike 
vectors, $\rho$  energy density and $p$ isotropic fluid pressure, 
${\bigtriangleup}p$ pressure anisotropy and the term involving $q$ 
represents heat flux.

\noindent We employ the comoving coordinates to write $u_i = \sqrt{g_{00}}
\delta^0_i$ and take $c_i = \sqrt{g_{11}} \delta^1_i$. The kinematic 
parameters; expansion, shear and acceleration for the metric (1) read as 
follows:
 
\begin{equation}
{\theta} = \frac{- \dot P r^2}{2(2r^2 + P) (r^2+ P)^{3/2}}, ~
\sigma^2 = \frac{2}{3} {\theta}^2, ~ \dot u_r = -\frac{r}
{r^2 + P}.
\end{equation}
 
\noindent Now applying the Einstein equation, we obtain \\
 
\begin{equation}
8 \pi \rho = \frac{2 r^2 + 3P}{(2r^2 + P)^2}
\end{equation}
 
\begin{equation}
8 \pi p_r = \frac{1}{2r^2 + P}
\end{equation}
 
\begin{equation}
8 \pi p_{\perp} = \frac{1}{2r^2 + P} + \frac{r^2}{4(2 r^2 + P)(r^2 + P)^2}
\left[2 \ddot P - \frac{(9r^2 + 5P) \dot P^2}{(2r^2 + P)(r^2 + P)} \right]
\end{equation}

\begin{equation}
8 \pi q = \frac{- \dot P r}{(2 r^2 + P)^{3/2}(r^2 + P)}.
\end{equation}
 
\noindent The pressure anisotropy $\bigtriangleup p = p_r - p_{\perp}$ is 
given by
 
\begin{equation}
8 \pi \bigtriangleup p = \frac{- r^2}{4(2 r^2 + P)(r^2 + P)^2}
\left[2 \ddot P - \frac{(9r^2 + 5P) \dot P^2}{(2r^2 + P)(r^2 + P)} \right].
\end{equation}

\noindent Now we choose for the time function
\be
P(t) = a^2 + b^2cos{\omega}t, ~a^2 > b^2. \\
\ee
\n This lends oscillatory behaviour to the model which oscillates between 
the two regular states. The oscillation period is $t = 
2\pi/{\omega}$, density is maximum at $t = (2n+1)\pi/{\omega}$ and it is 
minimum at $t = 2n\pi/{\omega}$ for an integer $n$. The model could have as 
low and as high density as one pleases by choosing large values for the 
parameters $a$ and $b$ with the former being as close to the latter 
from the above. The solution involves three parameters $a$, $b$ and 
$\omega$. Besides, when comparing with observational data, one needs a 
specification of the locale of observation,i.e. the time $t_0$ of 
observation and $r_0$ the radial coordinate of the observer. Since there is 
abundance of free 
parameters, it is therefore possible to coast the model as close to 
the observations as one pleases by suitale choice. \\

\n Above all the most interesting feature of the model is oscillatory 
behaviour indicating like the steady state cosmology no beginning and no 
end. As in QSSC [2], this model would also predict blue shifts, should 
they be observed in future, one need not necessarily have to invoke 
non-conservation of energy but instead an imperfect fluid distribution 
could as well do without violating GR and the usual energy conditions. 
This is quite remarkable and interesting feature of our model. \\

\n The pressure anisotropy and heat flux fall off as $r^{-4}$ and they vanish 
at the centre $r = 0$. It is obvious that expansion and heat flux have 
similar behaviour (note that in Ref. [1], there was a sign error which 
indicated opposite behaviour), vanishing at ${\omega}t = 0, \pi$ and 
attaining maximum at ${\omega}t = \pi/2, 3\pi/2$. That is in expanding 
phase heat flows out while the reverse happens for contracting phase. 
The pressure anisotropy could like density be made as small as one 
pleases and it changes sign at $t = (2n+1)/2 {\pi}{\omega}$. Acceleration 
vanishes at both ends, $r \rightarrow 0,\infty$ and is finite for all 
$t$ at a given $r$.\\ 
 
\noindent For the oscillatory choice (9), it can be easily checked that 
$\rho > p_r,
p_{\perp} > 0, (\rho + p_r)^2 - 4 q^2 > 0$ and $\rho - p_r - 2p_{\perp}
+ [(\rho + p_r)^2 - 4q^2]^{1/2} > 0$ always. This ensures that all the (weak, 
strong and dominant) energy conditions are satisfied. All the 
physical and kinematic parameters always remain regular and finite.. The  
metric is simple enough to see that it is causally stable and 
geodesically complete. \\ 

\noindent The present model gives a picture somewhat different from 
non-oscillating singularity-free models [1,3,4,6]. There the universe has 
a state of infinite dilution both in the infinite past and future and in 
between there is a state of maximum contraction. In our case $P(t)$ never 
becomes arbitrarily large and so there is no infinite dilution but 
periodically the physical variables oscillate between finite maxima and 
minima. However as $r \rightarrow \infty$, all the physical variables tend 
to vanish as in other non-singular models. This behaviour is demanded by 
the theorem [7] that in non-singular models, the space average of all 
the kinematic scalars and physical parameters must vanish (for a 
weaker theorem on the vanishing of space time averages [8].\\

\noindent We have thus shown that it is possible to have a truely 
oscillatory singularity free spherical model within GR without violating 
conservation of  energy and its usual conditions. Such a 
model would be consistent with blue shifts, should they be discovered in 
future. The model has a number of free parameters which can be suitably 
chosen to coast it arbitrarily close to the observations. One may not 
quite relish the introduction of an imperfect fluid but then we are not 
violating any stringent physical requirement. Abondoning the cosmological 
principle of isotropy and homogeneity requires hardly any apology but the 
introduction of a specially favoured centre of the universe may require 
some justification. However one cannot get the solution of the Einstein 
field equation without introducing some symmetry assumption. In any case 
the solution broadens our horizon about the potentialities of 
relativistic models.

\acknowledgments We wish to thank L. K. Patel and R. S. 
Tikekar and members of the Relativity and Cosmology Research Centre, 
Jadavpur University for helpful discussions. We also thank IUCAA 
for supporting visits of the authors that facilitated this work.

\end{document}